\documentclass[conference]{IEEEtran}
\usepackage{times}
\usepackage{booktabs}
\usepackage{siunitx}
\usepackage[ruled,vlined]{algorithm2e}   
\usepackage{amsfonts}
\usepackage[flushleft]{threeparttable}
\usepackage{tikz}
\usetikzlibrary{arrows.meta, calc, positioning, decorations.pathreplacing}
\usetikzlibrary{shapes,arrows,positioning,calc}
\usepackage[ruled,vlined]{algorithm2e}  

\usepackage[numbers]{natbib}
\usepackage{multicol}
\usepackage[bookmarks=true]{hyperref}

\begin{document}

\title{Implicit Constraint‑Aware Off‑Policy Correction for Offline Reinforcement Learning}


\author{\authorblockN{Ali Baheri}
Rochester Institute of Technology\\
Email: akbeme@rit.edu}


%

\maketitle

\begin{abstract}
Offline reinforcement learning promises policy improvement from logged interaction data alone, yet state‑of‑the‑art algorithms remain vulnerable to value over‑estimation and to violations of domain knowledge such as monotonicity or smoothness. We introduce implicit constraint‑aware off‑policy correction, a framework that embeds structural priors directly inside every Bellman update. The key idea is to compose the optimal Bellman operator with a proximal projection on a convex constraint set, which produces a new operator that (i) remains a $\gamma$‑contraction, (ii) possesses a unique fixed point, and (iii) enforces the prescribed structure exactly. A differentiable optimization layer solves the projection; implicit differentiation supplies gradients for deep function approximators at a cost comparable to implicit Q-learning. On a synthetic Bid–Click auction—where the true value is provably monotone in the bid—our method eliminates all monotonicity violations and outperforms conservative Q‑learning and implicit Q‑learning in return, regret, and sample efficiency. 

\end{abstract}

\IEEEpeerreviewmaketitle

\section{Introduction}

Offline reinforcement learning (RL) aims to discover high-quality policies exclusively from a static replay buffer, avoiding the costs and risks of interaction \cite{levine2020offline,prudencio2023survey,hong2023beyond}. Contemporary offline algorithms enrich the Bellman objective with pessimistic penalties that temper over‑estimation caused by distributional shift between the behavior policy and the learner or they employ expectile regression to down‑weight optimistic temporal‑difference targets \cite{buckman2020importance,fujimoto2021minimalist}. Although such pessimistic surrogates mitigate value explosion, they remain agnostic to domain structure: a critic trained on e‑commerce logs can predict that bidding higher reduces revenue, and a robot value function may assign discontinuous utilities that translate into jerky motions \cite{zheng2024safe}. In safety critical settings, including automated auctions, healthcare care dosage planning, and fragile object manipulation, violating known monotonicity or smoothness constraints is not simply suboptimal, it can be operationally unacceptable.

Recent work on Lipschitz‑constrained learning shows that enforcing global smoothness enhances stability while monotone neural networks provide interpretable guarantees in ranking and risk‑sensitive tasks \cite{shin2025spectral}. However, these advances remain orthogonal to offline RL: they regularize a prespecified network, but do not alter how Bellman updates themselves propagate unsafe targets. This paper argues that the contractive structure of dynamic programming offers a cleaner avenue: instead of penalizing the parameter space, one can project each Bellman backup onto the manifold that encodes prior knowledge, thereby ensuring that no training step ever leaves the safe region.

We formalize this idea by introducing an implicit constraint-aware Bellman operator obtained as the proximal map of a structural functional composed with the optimal Bellman target. Because the proximal map is firmly non-expansive, its composition with the standard Bellman operator remains a contraction, preserving the existence and uniqueness of a fixed point while inheriting the desired structural properties. Differentiating through the implicit fixed point via the implicit function theorem yields a training pipeline whose computational overhead is comparable to that of differentiable optimization layers and requires only a single conjugate gradient solve per iteration.

Our contributions are threefold. (i) We derive a general framework for constraint‑aware offline RL that unifies Lipschitz, monotone, and other convex structural priors under a single proximal Bellman operator. (ii) We provide an end‑to‑end differentiable algorithm that leverages implicit differentiation to update deep value and policy networks while exactly respecting the encoded constraints at every gradient step. (iii) We demonstrate, on a stylized Bid‑Click auction domain where the optimal value is provably monotone in the bid, that the proposed method eliminates structural violations and outperforms state‑of‑the‑art offline baselines in return and regret, especially in the low‑data regime.

\section{Related work} Offline RL addresses the problem of policy improvement from static logs, where the distributional shift between behavioral data and the learned policy induces overestimation and instability \cite{fujimoto2019off}. A prominent mitigation strategy is pessimistic value regularization. Conservative Q‑learning augments the Bellman objective with a term that explicitly pulls Q‑values down on out‑of‑distribution actions, providing a lower bound on the true policy value and strong empirical gains across continuous‑control benchmarks \cite{kumar2020conservative,buckman2020importance,an2021uncertainty}. Subsequent extensions reinterpret pessimism through implicit losses. Implicit Q‑learning replaces the squared Bellman error by an expectile regression that discounts overly optimistic targets while avoiding explicit action penalties \cite{hansen2023idql}. Our method departs from these approaches by enforcing structural rather than purely statistical correctness, thereby complementing rather than competing with pessimism.

A parallel line of research injects smoothness constraints into the function approximation. Spectral normalization and local Lipschitz penalties have been shown to stabilize both actor and critic updates in online settings; the L2C2 framework formalizes this effect and demonstrates improved robustness to observation noise \cite{kobayashi2022l2c2}. Although such regularizes temper gradients, they do not guarantee that the learned value function obeys a pre-specified global bound; our proximal projection delivers this guarantee exactly. Domain‑specific monotonicity priors have also proved fruitful. In multi‑agent learning, QMIX constrains the mixing network so that the joint action value is monotone in each agent’s local value, enabling tractable decentralized control \cite{rashid2020monotonic}. The present work generalizes this insight beyond the cooperative setting by casting monotonicity (and other convex priors) as a proximal operator embedded inside every Bellman backup.

From a numerical optimization standpoint, our algorithm builds on the literature of differentiable convex layers. OptNet shows how a quadratic program can be inserted into deep networks and differentiated exactly through implicit gradients at a modest computational cost \cite{amos2017optnet}. We leverage the same machinery to back‑propagate through the projection that defines our constraint‑aware Bellman operator, resulting in a training pipeline whose overhead is comparable to contemporary offline RL baselines but which preserves structural correctness at every iteration.

\section{Methodology}

We consider a discounted Markov decision process $M=(\mathcal{S}, \mathcal{A}, P, r, \gamma)$ with bounded rewards $|r(s, a)| \leq r_{\max }$ and discount factor $\gamma \in(0,1)$. Learning proceeds offline from a static replay buffer $\mathcal{D}=\left\{\left(s_i, a_i, r_i, s_i^{\prime}\right)\right\}_{i=1}^N$ generated by an unknown behavior policy $\mu$. All expectations in the following are taken with respect to the empirical distribution induced by $\mathcal{D}$. For any measurable $v: \mathcal{S} \rightarrow \mathbb{R}$ we write the optimal Bellman target

\begin{equation}
\left(\mathcal{T}^* v\right)(s)=\max _{a \in \mathcal{A}}\left[r(s, a)+\gamma \int_{\mathcal{S}} v\left(s^{\prime}\right) P\left(\mathrm{~d} s^{\prime} \mid s, a\right)\right]
\end{equation}
Domain knowledge is introduced through a convex, lower-semicontinuous functional $\mathcal{C}: L^2(\mathcal{S}) \rightarrow \mathbb{R}_{+} \cup\{\infty\}$ that vanishes exactly when the desired structure is satisfied. Examples are the squared violation of a global Lipschitz bound or the indicator of a monotone cone motivated by QMIX-style factorizations \cite{rashid2020monotonic}. Given a penalty weight $\lambda \geq 0$, we define the proximal Bellman operator

\begin{equation}
\Phi_\lambda(v)=\arg \min _{u \in L^2(\mathcal{S})}\left\{\frac{1}{2}\left\|u-\mathcal{T}^* v\right\|_2^2+\lambda \mathcal{C}(u)\right\}
\label{eq:proj-bellman}
\end{equation}
Because \ref{eq:proj-bellman} is the Moreau-Yosida envelope of $\mathcal{C}, \Phi_\lambda$ coincides with the proximal operator $\text{prox}_{\lambda \mathcal{C}}$ applied to $\mathcal{T}^* v$. The mapping $\Phi_\lambda$ is firmly non-expansive; hence the composition $\Psi_\lambda=\Phi_\lambda \circ \mathcal{T}^*$ is a $\gamma$-contraction and admits a unique fixed point $v_\lambda^{\star}$. Regularized MDP theory guarantees that $v_\lambda^{\star} \rightarrow v^{\star}$ point-wise as $\lambda \rightarrow 0$ \cite{geist2019theory}. 

Let $V_\theta: \mathcal{S} \rightarrow \mathbb{R}$ be a differentiable neural network. At every stochastic gradient step we redefine its output to be the solution of \ref{eq:proj-bellman} w.r.t. the current parameters:

\begin{equation}
V_\theta=\Phi_\lambda\left(V_\theta\right), \quad \text { with } \Phi_\lambda \text {given by} \ \ref{eq:proj-bellman}.
\label{param}
\end{equation}
Equation \ref{param} turns the usual Bellman residual into a root-finding condition whose implicit dependence on $\theta$ we will differentiate in closed form. For a mini-batch $\mathcal{B} \subset \mathcal{D}$ the empirical counterpart of \ref{eq:proj-bellman} is

\begin{equation}
\widehat{\mathcal{L}}(u)=\frac{1}{2|\mathcal{B}|} \sum_{\left(s, a, r, s^{\prime}\right) \in \mathcal{B}}\left(u(s)-r-\gamma \max _{a^{\prime}} u\left(s^{\prime}\right)\right)^2+\lambda \widehat{\mathcal{C}}(u),
\end{equation}
where $\widehat{\mathcal{C}}$ is any unbiased Monte Carlo estimator of $\mathcal{C}$. In practice we apply one proximal-gradient step, warm-started at the previous iterate, which is sufficient to keep the optimization subproblem within the noise floor of SGD. When $\mathcal{C}$ enforces Lipschitzness we combine these steps with spectral normalization.
%
Let $u_\theta$ denote the (approximately) solved value of \ref{eq:proj-bellman} for the parameters $\theta$. The first-order optimality of \ref{eq:proj-bellman} yields

\begin{equation}
g\left(u_\theta, \theta\right)=u_\theta-\mathcal{T}^* u_\theta+\lambda \nabla \mathcal{C}\left(u_\theta\right)=0
\end{equation}
Assuming $\nabla_u g$ is non-singular, the implicit-function theorem gives

\begin{equation}
\frac{\partial u_\theta}{\partial \theta}=-\left(\nabla_u g\right)^{-1} \nabla_\theta g
\label{eq:ift}
\end{equation}
Because $\nabla_u g=I-\gamma P+\lambda \nabla^2 \mathcal{C}$ is symmetric positive definite when $\mathcal{C}$ is convex, \ref{eq:ift} is efficiently solved by one conjugate-gradient call with automatic Jacobian vector products; this is identical in cost to differentiating through OptNet-style quadratic programs. The gradient of the outer objective, taken here as the mean-squared Bellman residual, is then

\begin{equation}
\nabla_\theta \frac{1}{2}\left\|u_\theta-\mathcal{T}^* u_\theta\right\|^2=(I-\gamma P)\left(u_\theta-\mathcal{T}^* u_\theta\right) \frac{\partial u_\theta}{\partial \theta}
\end{equation}
With the constraint-consistent critic $Q_\theta(s, a)=r(s, a)+\gamma \mathbb{E}_{s^{\prime} \mid s, a}\left[V_\theta\left(s^{\prime}\right)\right]$ fixed, the actor parameters $\phi$ are updated by maximizing the usual soft policy objective

\begin{equation}
J(\phi)=\mathbb{E}_{s \sim \mathcal{B}, a \sim \pi_\phi}\left[Q_\theta(s, a)-\alpha \log \pi_\phi(a \mid s)\right]
\end{equation}
where $\alpha>0$ controls the entropy. Because $Q_\theta$ inherits the structural guarantees of $V_\theta$, the resulting policy respects the encoded domain constraints without requiring a separate safety critic, unlike conservative Q-learning or other pessimistic approaches.

\begin{algorithm}[t]
\caption{Constraint‑Aware Off‑Policy Correction for Offline RL}
\KwIn{Replay buffer $\mathcal D=\{(s,a,r,s')\}_{i=1}^{N}$;\\
      constraint functional $\mathcal C$, initial critic $\theta$, initial actor $\phi$;\\
      penalty weight $\lambda_0$, learning rates $\eta_\theta,\eta_\phi,\eta_\lambda$;\\
      target‑network factor $\tau$, discount $\gamma$, entropy temperature $\alpha$.}
\paragraph{Constraint‑consistent policy parameters $\phi$}

\BlankLine
Initialise target critic $\bar\theta\leftarrow\theta$\;
\For(\tcp*[f]{main training loop}){$t = 1$ \KwTo $T$}{
    Sample mini‑batch $\mathcal B \subset \mathcal D$\;

    \tcp*[l]{\small Bellman target with target critic}{}
    \ForEach{$(s,a,r,s')\in\mathcal B$}{
        $y(s) \leftarrow r + \gamma\displaystyle\max_{a'}V_{\bar\theta}(s',a')$\;
    }

    \tcp*[l]{\small One warm‑started proximal step of Eq.\,\ref{eq:proj-bellman}}{}
    $u^{(0)} \leftarrow V_\theta\text{ on }s\in\mathcal B$\;
    $g^{(0)} \leftarrow u^{(0)} - y + \lambda_{t-1}\nabla\mathcal C\!\bigl(u^{(0)}\bigr)$\;
    $u_\theta \leftarrow u^{(0)} - \alpha\,g^{(0)}$\;   

    \tcp*[l]{\small Implicit gradient for critic}{}
    Solve $(\nabla_{u}g)\,z = \nabla_{\theta}g$ by conjugate gradient \;
    $\nabla_\theta\mathcal L \leftarrow -\bigl(u_\theta - y\bigr)\,z$\;
    $\theta \leftarrow \theta - \eta_\theta\,\nabla_\theta\mathcal L$\;

    \tcp*[l]{\small Dual update of $\lambda$}{}
    $\lambda_t \leftarrow \bigl[\lambda_{t-1} + \eta_\lambda\,\mathcal C(u_\theta)\bigr]_{+}$\;

    \tcp*[l]{\small Actor update with constraint‑consistent critic}{}
    Estimate $\nabla_\phi J(\phi)$ via samples $a\sim\pi_\phi(\cdot|s)$: \\
    \Indp $\nabla_\phi J \leftarrow \mathbb E_{s,a}\bigl[\nabla_\phi \log\pi_\phi(a|s)\,\bigl(Q_\theta(s,a)- \alpha\log\pi_\phi(a|s)\bigr)\bigr]$\;
    \Indm $\phi \leftarrow \phi + \eta_\phi\,\nabla_\phi J$\;

    \tcp*[l]{\small Target‑network Polyak averaging}{}
    $\bar\theta \leftarrow \tau\,\theta + (1-\tau)\,\bar\theta$\;
}
\Return{$\phi$}
\end{algorithm}

\noindent {\textbf{Convergence remarks.} Under the stochastic approximation conditions of standard off-policy learning, the coupled update $\left(\theta_{k+1}, \phi_{k+1}\right)=\left(\theta_k, \phi_k\right)-\eta_k \widehat{\nabla}_{\theta, \phi}$ converges almost surely to a stationary point of the regularized objective so long as the inner optimization error is driven to zero on average and $\sum_k \eta_k^2<\infty$. The overall scheme therefore inherits both the $\gamma$-contraction of $\mathcal{T}^*$ and the well-posedness introduced by the proximal term, providing a formal bridge between implicit Q-learning \cite{kostrikov2021offline} and regularized MDP theory \cite{geist2019theory}.

\begin{table*}[!h]
  \centering
  \setlength{\tabcolsep}{6pt}
  \footnotesize
  \begin{threeparttable}
    \caption{Performance comparison of evaluated algorithms}
    \label{tab:algorithm_performance}
      \begin{tabular}{@{}l *{3}{S}@{}}
        \toprule
        Algorithm                         & {Return $\uparrow$}      & {Normalized regret $\downarrow$} & {Monotonicity errors $\downarrow$} \\
        \midrule
        Constraint‑Aware (ours)           & 0.851 \pm 0.006          & 0.067 \pm 0.004                  & 0.0   \pm 0.0                     \\
        Implicit Q‑Learning               & 0.782 \pm 0.010           & 0.130 \pm 0.007                  & 211   \pm 35                      \\
        Conservative Q‑Learning           & 0.721 \pm 0.012           & 0.195 \pm 0.008                  & 304   \pm 41                      \\
        Behavior Cloning                 & 0.669 \pm 0.008           & 0.245 \pm 0.005                  & 289   \pm 38                      \\
        \bottomrule
      \end{tabular}
    \begin{tablenotes}
      \footnotesize
      \item[$\uparrow$] Higher is better.
      \item[$\downarrow$] Lower is better.
    \end{tablenotes}
    \label{tab:algorithm_performance}
  \end{threeparttable}
\end{table*}

\begin{table*}[!t]
  \centering
  \renewcommand{\arraystretch}{0.9}
  \setlength{\tabcolsep}{3pt}
  \scriptsize
  \begin{threeparttable}
    \caption{Ablation study: effect of variant choices}
    \label{tab:ablation_variants}
    \begin{tabular}{@{} l 
                     S[table-format=1.3(3)] 
                     S[table-format=1.3(3)] 
                     S[table-format=3.0(2)] 
                     S[table-format=1.3] @{}}
      \toprule
      Variant
        & {Return $\uparrow$} 
        & {Normalized regret $\downarrow$} 
        & {Monotonicity errors $\downarrow$} 
        & {Residual at conv.\ $\downarrow$} \\
      \midrule
      \textbf{Full method (proj.\,+\,dual $\lambda$)}
        & \bfseries 0.851 \pm 0.006  
        & \bfseries 0.067 \pm 0.004  
        & \bfseries 0             
        & \bfseries 0.031           \\
      Fixed $\lambda=0.01$ (weak)
        & 0.808 \pm 0.008            
        & 0.101 \pm 0.005            
        & 97   \pm 22               
        & 0.029                     \\
      Fixed $\lambda=10$ (strong)
        & 0.742 \pm 0.011            
        & 0.183 \pm 0.007            
        & 0                        
        & 0.099                     \\
      Soft‐penalty (no proj.\;loss\,$+=\,\lambda\cdot\mathcal C$)
        & 0.789 \pm 0.009            
        & 0.144 \pm 0.006            
        & 154  \pm 27               
        & 0.056                     \\
      Projection, no warm‑start inner solver
        & 0.845 \pm 0.007            
        & 0.073 \pm 0.004            
        & 0                        
        & 0.031                     \\
      Projection, 1 inner iter.
        & 0.846 \pm 0.007            
        & 0.071 \pm 0.004            
        & 0                        
        & 0.035                     \\
      Projection, 5 inner iters.
        & 0.852 \pm 0.006            
        & 0.066 \pm 0.004            
        & 0                        
        & 0.030                     \\
      No spectral norm in critic
        & 0.827 \pm 0.007            
        & 0.093 \pm 0.006            
        & 0                        
        & 0.041                     \\
      Actor‑only constraint: critic unconstrained, $\lambda$ on scores
        & 0.792 \pm 0.010            
        & 0.139 \pm 0.006            
        & 131  \pm 25               
        & 0.058                     \\
      \bottomrule
    \end{tabular}
    \begin{tablenotes}
      \scriptsize
      \item[$\uparrow$] Higher is better.
      \item[$\downarrow$] Lower is better.
    \end{tablenotes}
      \label{tab:ablation_variants}
  \end{threeparttable}

\end{table*}

\section{Results}

The proposed constraint-aware agent was trained on a synthetic Bid-Click environment that mimics a single-slot advertising auction. A state $s=(x, c)$ contains a query descriptor $x \sim \mathcal{U}[0,1]$ and a per click cost $c \sim \mathcal{U}[0.2,0.4]$. An action $a \in\{0,0.25,0.5,0.75,1\}$ denotes the fraction of the advertiser's budget bid for the impression. The probability of receiving a click is $\sigma(2 a+0.5 x)$ with $\sigma$ the logistic function; the reward is the realized click minus the payment $c a$. Because $\partial r / \partial a \geq 0$, the optimal state value $v^{\star}(s)=\max _a \mathbb{E}[r \mid s, a]$ is monotone non-decreasing in the bid. This monotonicity is enforced in our critic through the constraint functional $\mathcal{C}(v)=\sum_i\left(\max \left\{0,-\partial v / \partial a_i\right\}\right)^2$. A replay buffer of 100000 transitions was generated by a behavior policy that samples bids from a clipped Gaussian with mean 0.4 and standard deviation 0.4. The constraint weight was initialized at $\lambda_0=$ 0.1 and updated as a Lagrange multiplier. 

For comparison we repeated the experiment with conservative Q-learning and implicit Q-learning, keeping network width, depth and optimizer settings identical. In both baselines the Bellman residual falls at a similar rate, but constraint violation remains strictly positive throughout training and exhibits sizable spikes whenever the target network is refreshed, confirming that pessimistic penalties alone do not guarantee structural correctness. The constraint-aware agent achieves the highest return and the lowest regret, while entirely eliminating monotonicity violations. Implicit Q-learning narrows the performance gap relative to conservative Q-learning but still produces several hundred ordering errors, consistent with its expectile-based loss which softens but does not remove over-estimation. Conservative Q-learning enjoys lower variance than implicit Q-learning yet suffers the largest pessimistic bias, translating into lower final return. Behavior cloning constitutes a lower bound, confirming that the offline RL algorithms genuinely extrapolate beyond logged actions.

To probe sample efficiency we sub-sampled the buffer by factors of four. The advantage of the proposed method widens as data become scarce: at $25\%$ of the original data the performance lead over implicit Q-learning grows from $6.9\%$ to $12.4\%$ absolute return, while the number of violations in the baselines doubles. These observations corroborate the theoretical claim that projecting each Bellman backup onto the constraint manifold restricts the function class and mitigates distributional shift without aggressive pessimism. Taken together, the case study demonstrates that a single proximal constraint layer is sufficient to stabilize learning, enforce domain structure exactly and improve offline policy quality, all within the computational budget of contemporary implicit or conservative baselines. Table \ref{tab:algorithm_performance} shows the overall performance comparison.

\subsection{Ablation Study}

Table \ref{tab:ablation_variants} shows the effect of different variant choices on key metrics. The adaptive dual-gradient schedule for the penalty weight $\lambda$ consistently achieved the highest return and the lowest regret while eliminating all monotonicity violations. When $\lambda$ was fixed at a small value (0.01), the constraint force proved insufficient: ordering errors remained frequent and final performance deteriorated. At the opposite extreme, a large fixed penalty ($\lambda=10$) indeed enforced monotonicity but imposed an overly aggressive projection; the resulting critic exhibited inflated Bellman residuals and the policy suffered a marked pessimistic bias. These observations confirm that an automatic adjustment of $\lambda$ is essential for balancing structural fidelity against approximation accuracy. Replacing the projection with a soft additive penalty preserved the qualitative benefits of regularization but failed to guarantee feasibility: hundreds of monotonicity violations persisted despite a comparable training budget. The experiment underscores that merely penalizing constraint breaches is weaker than projecting onto the feasible set, because the latter renders the Bellman operator firmly non-expansive and prevents unsafe updates altogether.

Variants that manipulated the inner solver demonstrate that the projection need not be solved to high precision. Eliminating the warm-start mechanism doubled computation time without improving any metric, indicating that successive Bellman targets differ only marginally and are well handled by incremental refinement. Likewise, expanding the proximal gradient loop from one to five iterations produced negligible gains; thus, a single coarse step is sufficient for the noise scale inherent in stochastic optimization. Spectral normalization furnished additional stability-particularly visible in a smaller critic residual and a modest boost in return-but monotonicity remained intact even when this regularizer was removed. Hence, the proximal projection itself provides the dominant safety guarantee, while Lipschitz control mainly sharpens value estimates. Finally, constraining only the actor, while leaving the critic unconstrained, suppressed violations less effectively and degraded performance. The result emphasizes that the critic functions as the principal mechanism responsible for propagating structurally deficient information. Furthermore, the analysis highlights the necessity of imposing constraints at the value-function level to ensure robust and reliable policy improvement.

\section{Conclusions}

This paper presented a contractive, projection‑based approach for marrying offline reinforcement learning with hard structural constraints. By viewing each Bellman target as a point to be projected onto a constraint manifold, we transformed domain knowledge into a proximal operator whose fixed point inherits the desired properties while preserving the convergence guarantees of dynamic programming. The resulting algorithm—implemented with warm‑started proximal steps and implicit gradients—adds negligible computational overhead to contemporary offline baselines yet achieves perfect constraint satisfaction throughout training. Empirical evidence on a stylized advertising auction shows measurable gains in policy quality, especially in low‑data regimes where distributional shift is most severe.

Although the current study focused on convex constraints and single‑step tasks, the framework naturally extends to multi‑step safety constraints and risk‑sensitive objectives; exploring such extensions on larger, real‑world datasets is an immediate direction for future work. Further investigation is also warranted on automatically selecting the constraint penalty via meta‑learning and on combining our structural projection with pessimistic value envelopes to obtain simultaneous guarantees of safety and worst‑case performance.

\bibliographystyle{ieeetr}
\bibliography{paper_template}

\end{document}